\documentclass{kluwer}    

\usepackage{graphicx}

\newdisplay{guess}{Conjecture}

\begin{document}
\begin{article}

\begin{opening}         

\title{Chemo-dynamical Evolution of the ISM in Galaxies} 

\author{Stefan \surname{Harfst}\thanks{harfst@astrophysik.uni-kiel.de}}
\author{Gerhard \surname{Hensler}}
\author{Christian \surname{Theis}}
\institute{Institute for Theoretical Physics and Astrophysics, University Kiel, Germany}

\runningauthor{S. Harfst et al.}
\runningtitle{Chemo-dynamical Evolution of the ISM in Galaxies}

\begin{abstract}
Chemo-dynamical models have been introduced in the late eighties and
are a generally accepted tool for understanding galaxy evolution. They
have been successfully applied to one-dimensional problems, e.g.\ the
evolution of non-rotating galaxies, and two-dimensional problems,
e.g.\ the evolution of disk galaxies. Recently, also three-dimensional
chemo-dynamical models have become available. In these models the
dynamics of different components, i.e. dark matter, stars and a
multi-phase interstellar medium, are treated in a self-consistent way
and several processes allow for an exchange of matter, energy and
momentum between the components or different gas phases.  Some results
of chemo-dynamical models and their comparison with observations of
chemical abundances or star formation histories will be reviewed.
\end{abstract}


\end{opening}           

\section{Introduction}  
\label{harfst_sec_intro}

Disk galaxies consist of complex structures with at least three main
components: bulge, halo and disk. These differ in their fraction of
ionised, atomic and molecular gas, their dust content and their
stellar populations. Since the Milky Way Galaxy (MWG) is the best
studied disk galaxy, the observations of its stars and its
interstellar medium (ISM) can be used to test galactic evolutionary
models in order to answer the addressed fundamental questions
concerning the initial conditions, the formation and the evolution of
the MWG and of disk galaxies in general, e.g.: When, how and on what
time-scales did the galactic components form? Was there any connection
between them? Which external influences have affected the structure? 
Is the solar neighbourhood representative for an "average disk"? These
questions are directly related to observational facts, such as the
lack of metal-poor G-dwarfs (the well-known G-dwarf problem) or the
different observed effective yields in bulge, halo and disk of the MWG
\cite{Pagel87}. In general, these and other observational indicators,
like star formation (SF) and supernova (SN) rates or colour
measurements place constraints on galactic models and should be
explained in a global scenario.

Another class of objects are dwarf galaxies (DGs) which differ in
their structural and chemical properties from those of giant
galaxies. Because of their low gravitational energies DGs are greatly
exposed to energetic influences from processes like stellar winds, SN
or even stellar radiation. In addition, low-mass galaxies seem to form
at all cosmological epochs and by different processes, which make them
ideal laboratories to investigate many of the astrophysical processes
relevant for galaxy evolution. Again, the observations of DGs, like SF
rates or metal abundances and abundance ratios, should be explained in
a global model of galactic evolution.

At present, two major and basically different strategies for modelling
galaxy evolution are followed: firstly, dynamical investigations which
include hydro-dynamical simulations of isolated galaxy evolution and
of proto-galactic interactions reaching from cosmological perturbation
scales to direct mergers; and, on the other hand, studies which
neglect any dynamical effects but consider either the whole galaxy or
particular regions and describe the temporal evolution of mass
fractions and element abundances in detail. Simulating different
galactic regions, e.g. by a closed-box model, however, presupposes
that these regions are neither energetically nor dynamically
coupled. Thus, in this picture a galaxy is only the sum of different
isolated subsystems without any connection, despite the initial
conditions. Even multi-zone models which include gas exchange between
different galactic regions (e.g. \opencite{Ferrini94}) are devoid of
self-consistent gas dynamics.

Chemo-dynamical models are a different approach to modelling galactic
evolution treating self-consistently all important dynamical and
energetic processes as well as their dependence on the
metallicity. For this, a multi-phase description of the ISM with the
inclusion of star-gas interactions is essential. Such models have been
applied successfully to spherical symmetric systems (e.g.\
\opencite{TBH92}) and to axisymmetric systems (e.g.\
\opencite{SHT97}, hereafter SHT). More recently with increasing
computational power, also 3d chemo-dynamical models have become
available (e.g.\ \opencite{SG02}). The chemo-dynamical treatment is
described in more detail in Sec. \ref{harfst_sec_cdt} and in
Sec. \ref{harfst_sec_res2d} some results of two-dimensional models are
reviewed.

\section{Chemo-dynamical treatment}
\label{harfst_sec_cdt}

\begin{figure}[t]
\centerline{\includegraphics[width=\textwidth]{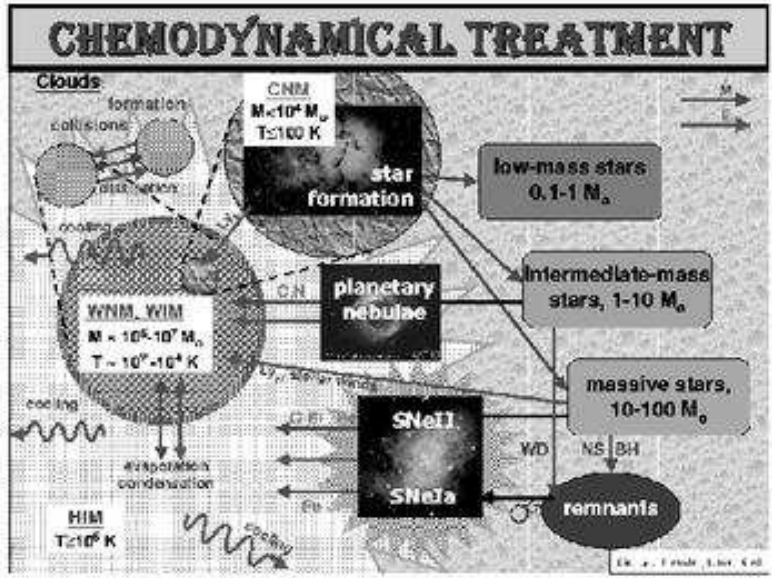}}
\caption{
A schematic sketch of the chemo-dynamical treatment. The different
gaseous and stellar components are connected via mass, momentum and
energy exchange. The important interaction processes are shown in this
diagram.}
\label{harfst_fig1}
\end{figure}

To approach global models of galaxy evolution which can
achieve the structural differences and details, an appropriate
treatment of the dynamics of stellar and gaseous components is
essential. In addition, at least the following processes should be
taken into account: SF, SNe, heating, cooling, stellar mass loss,
condensation and evaporation. This includes the treatment of the
multi-phase character of the ISM as well as the star-gas interactions
and phase transitions. Since gas and stars evolve dynamically, and
because several processes both depend on their metallicities but also
influence the element abundances in each component, these models are
called chemo-dynamical. The network of chemo-dynamical processes is
sketched in Fig. \ref{harfst_fig1} where one can see how the different
gaseous and stellar components are connected via mass, momentum and
energy exchange. The important processes are: mass from the cloudy
medium is transformed into stars by SF; the high and intermediate mass
stars return a fraction of this mass, enriched with metals, by SNe and
PNe to the ISM; the ISM is also heated by stellar radiation and
feedback processes; remnants and low mass stars build the lock-up mass
that is no longer available for the galactic cycle of matter; the
cloudy medium (CM) and the hot intercloud medium (ICM) are mixed by
condensation and evaporation; energy is dissipated by radiative
cooling and by cloud-cloud collisions.

As can be shown and must be emphasized, however, the number of free
parameters in the chemo-dynamical scheme is small, because they are either
theoretically evaluated (like e.g. condensation and evaporation) or
empirically determined (e.g. like stellar winds), or because they
force self-regulation in a way that is independent of the
parameterisation. The latter has been shown for SF by
\citeauthor{KTH95} (\citeyear{KTH95}, \citeyear{KTH98}). The only free
parameters which cannot approach particular "equilibrium values" due
to the absence of feedback are the stellar initial mass function
(IMF), the momentum transfer by drag and the initial conditions,
although the initial density distribution and gas-phase fraction are
basically not affecting the model evolution. For a more comprehensive
description of the chemo-dynamical treatment the interested reader is
referred to \inlinecite{TBH92} and SHT.

\section{Results from chemo-dynamical models}
\label{harfst_sec_res2d}

\subsection{The Milky Way's chemo-dynamical evolution}
\label{harfst_sec_resmwg}

\begin{figure}[t]
\centerline{\includegraphics[width=\textwidth]{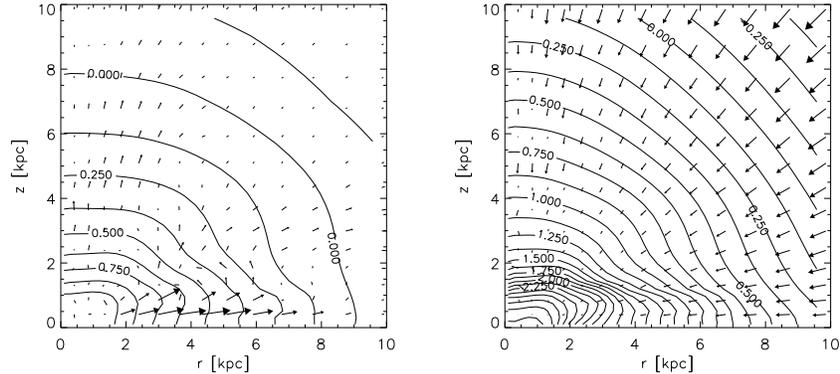}}
\caption{
Surface densities and velocities of ICM (left) and CM (right) 7\,Gyr
after the onset of the galactic collapse. The ICM in the galactic
plane has velocities up to $230\,{\rm km}\,{\rm s}^{-1}$,
while the velocities of the infalling CM are less than $60\,{\rm
km}\,{\rm s}^{-1}$. (taken from SHT)}
\label{harfst_fig2}
\end{figure}

\begin{figure}[t]
\centerline{\includegraphics[width=\textwidth]{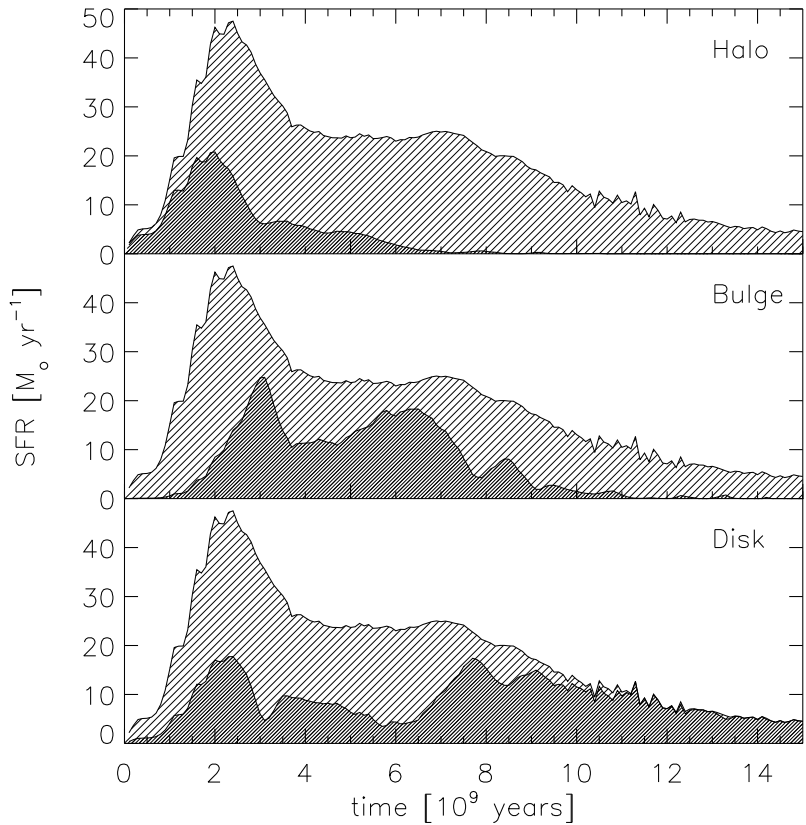}}
\caption{
The evolution of SF rate in different galactic regions. The upper
curve always shows the total SF rate, while the lower curve indicates
the SF rate in halo, bulge and disk. (taken from SHT)}
\label{harfst_fig3}
\end{figure}

The model starts from an isolated spheroidal, rotating but purely
gaseous cloud with a mass of $3.7 \cdot 10^{11}\,{\rm M}_{\odot}$,
a radius of 50\,kpc, and an angular momentum of about $2 \cdot
10^7\,{\rm M}_{\odot}\,{\rm pc}^2\,{\rm Myr}^{-1}$,
corresponding to a spin parameter $\lambda = 0.05$. It is assumed that
the protogalaxy consists initially of CM and ICM with a density
distribution of Plummer-Kuzmin-type \cite{Satoh80} with 10\,kpc scale
length. The initial CM/ICM mass division (99\%/1\%) does not affect
the later collapse, because the onset of SF determines the physical
state within less than $10^7$ years.  Since almost excellent agreement
of the chemo-dynamical model after 15\,Gyr is found with the presently observed
structure of the MWG, it may be safe to assume that also the
evolutionary behaviour of different properties under consideration can
be reliable deduced. The 15\,Gyr old model model is e.g. able to
reproduce the different metallicity distributions and effective yields
of the halo, the bulge and in the solar vicinity (see Fig. 7 in SHT),
respectively, and also to solve the G-dwarf problem as an effect of
large-scale dynamics of the metal-enriched ICM and its condensation
and metal pollution of the CM (Fig. \ref{harfst_fig2}). Furthermore,
the radial abundance gradients (see Fig. 6 in SHT), abundance ratios
like O/Fe versus Fe/H (see Fig. 9 in SHT), mass fractions of the
components, SF rate in the disk, etc. fit the observations strikingly.
Also interestingly, one can trace the temporal run e.g. of SN and PN
rates (see Fig 8. in SHT) and of the SF rates in different MWG regions
(Fig. \ref{harfst_fig3}) which allow the dating of their formation,
showing a delay of disk formation with respect to the halo extinction
by almost 4 Gyr and a bulge evolution extended over 5 Gyr with a least
two SF episodes.

\subsection{A Dwarf Galaxy model}
\label{harfst_sec_resdg}

\begin{figure}[t]
\centerline{\includegraphics[width=\textwidth]{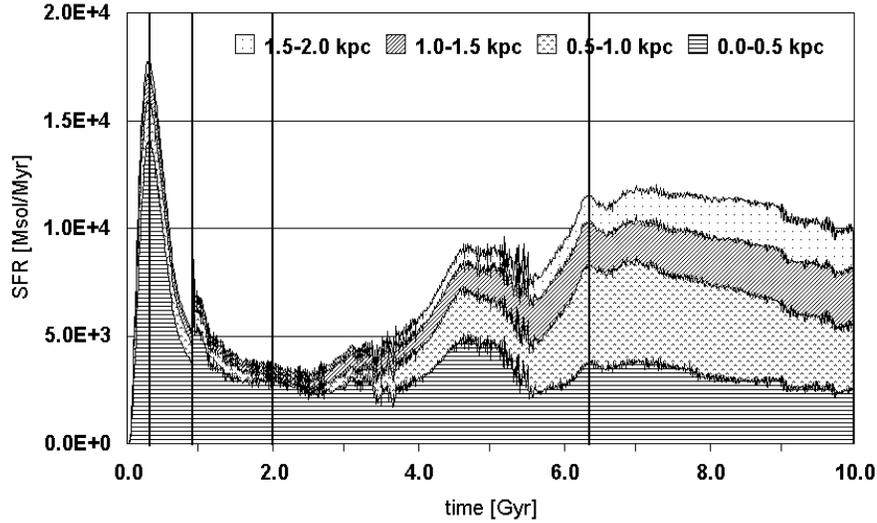}}
\caption{
Star formation history of the $10^9\,{\rm M}_\odot$ chemo-dynamical
dIrr model in units of ${\rm M}_\odot\,{\rm Myr}^{-1}$ for different
radial zones in the equatorial plane. The absolute values correspond
to the differences between two curves. The vertical lines divide the
different evolutionary phases as described in the text. (taken from RH)}
\label{harfst_fig4}
\end{figure}

\begin{figure}[t]
\centerline{\includegraphics[width=\textwidth]{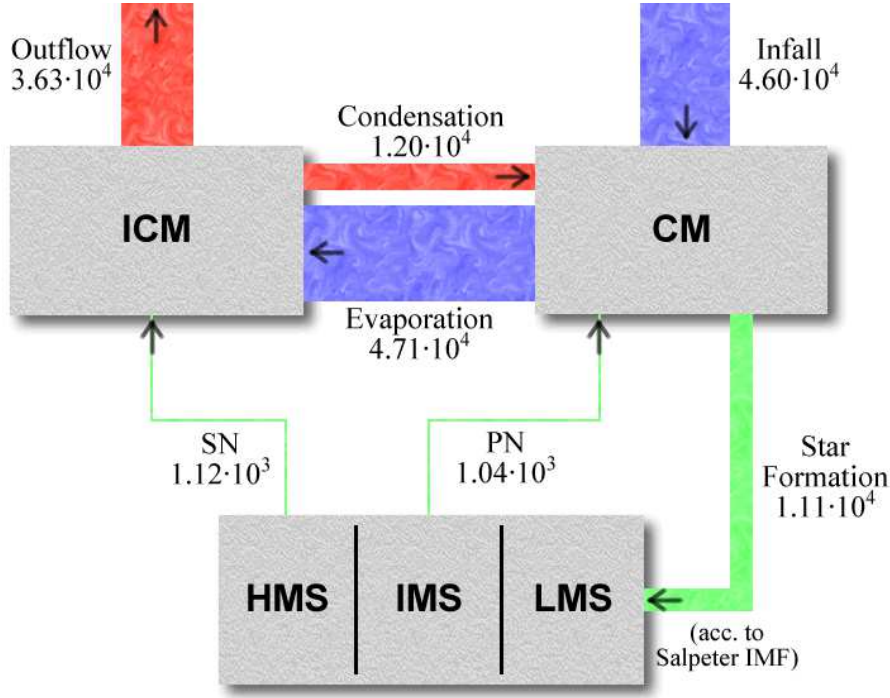}}
\caption{
Flow of matter between the components in the chemo-dynamical dIrr
model. The diagram shows the temporal average over the interval 6.2 to
10.0\, Gyr (the irregular phase, see text) for a radius of 2.0\,kpc
and a z-height of 1.0\,kpc, i.e. the whole visible galaxy. All numbers
are in units of  ${\rm M}_\odot\,{\rm Myr}^{-1}$. (taken from RH)}
\label{harfst_fig5}
\end{figure}

In this model by \citeauthor{RH00} (\citeyear{RH00}, hereafter RH) the
initial baryonic mass is about $10^9\,{\rm M}_{\odot}$ embedded in a
static dark matter halo according to \inlinecite{Burkert95} with a mass
of $10^{10}\,{\rm M}_{\odot}$. The numerical grid size is 20\,kpc x
20\,kpc, and the model is aimed to represent a dwarf irregular galaxy
(dIrr). In Fig. \ref{harfst_fig4} the SF history is plotted and five
distinct dynamical phases of evolution, that can be distinguished from
a kinematical analysis of the model, are indicated by vertical
lines. In the collapse phase (0 -- 0.3\,Gyr) the proto-galactic gas
distribution cools and collapses with a net gas infall rate of $3.2
\cdot 10^{-1}\,{\rm M}_{\odot}\,{\rm yr}^{-1}$ leading to a central
density increase by a factor of 100. Thus the SF rate rises steeply
according to its quadratic dependence on the CM-density to its maximum
value of $1.8 \cdot 10^{-2}\,{\rm M}_{\odot}\,{\rm yr}^{-1}$. 

After passing a post-collapse, a transitional and a turbulent phase
(see RH for details) the galaxy reaches its final irregular phase
after 6.2\,Gyr, when a global quasi-equilibrium seems to be
established. At this stage the disk is balancing itself in the sense
of keeping the gas mass constant. This can be seen in
Fig. \ref{harfst_fig5} where the mass flows between the different
components averaged between 6.2 - 10.0\,Gyr over a cylinder with $r =
2\,{\rm kpc}$ and $z = 1\,{\rm kpc}$ are analyzed. From the mass flow
rates one can distinguish between an outer and an inner cycle. The
outer one is produced by infall of CM. The CM-reservoir in the galaxy
is consumed by means of SF. The SF rate amounts to about 24\% of the
infall rate and 10\% of this matter is almost instantaneously reejected
by high mass stars (HMS). The inner cycle represents SF and stellar
evolution and, therefore, contains different time-scales.

The production of a minor fraction of hot ICM by SNe II leads to
evaporation of remaining CM and escapes from the galactic body as
outflow as a consequence of its high energy content. Significantly,
almost 80\% of the infall is immediately converted into outflow by
only 3\% of gas that has gone through the stellar cycle and is puffed
up by stellar energy release. The full evaporation rate can exceed the
infall rate because shell sweep-up leads to fragmentation and cloud
formation in addition to a small amount of condensation. It should be
emphasized that even though a large amount of the outflowing gas is
gravitationally unbound and leaves the galactic body, the metals
produced in HMS are for the most part kept in the outer gas flow cycle
by mixing ICM with continuously infalling clouds. As a result of this
mechanism only a few percent of the metals leave the gravitational
field of the galaxy with the outflowing ICM. While this mixing itself
happens on a time-scale of about 20\,Myr, the complete cycle of metal
enrichment takes almost 1\,Gyr because of the low infall velocity at
later evolutionary stages. In contrary, the inner cycle leads to an
efficient self-enrichment of SF regions within 10 Myr (see also
\opencite{SDK98}). The outer enrichment time-scale would be reduced
significantly, however, in a scenario of a rapidly infalling
intergalactic gas cloud.

\section{Summary and outlook}

Chemo-dynamical models of the MWG and dIrrs have been
reviewed. Results demonstrate both the agreement of one single global
model with numerous detailed and regional observations as well as with
the global structures. The necessity of the full but complex
chemo-dynamical treatment of the galaxy components is
justified. Several observational features like abundance distributions
and ratios can only be understood in a self-consistent global
evolutionary scenario if one is taking the dynamics of the components
and their relevant interaction processes into account. It has been
shown that a large-scale coupling of different galactic regions by
dynamical effects as well as the small-scale mixing between the gas
phases due to condensation/evaporation affect the observational
signatures. As long as the long-range streaming with inherent
small-scale interactions under the inclusion of a two-phase ISM with
their small-scale spatial resolution cannot be treated in other
numerical simulations, e.g. in present SPH models, they miss
self-consistency and cannot reproduce the observations in their global
extent.  

One has to emphasize that a reliable chemo-dynamical model has to
reproduce the complete set of observational features on different
galactic scales, i.e. for all existing components and observed
variables as the whole. It is not sufficient for an understanding of
the global galactic evolution to fit single particular observed
properties. The time dependence of element abundances and their ratios
can serve as reliable diagnostic tools of galaxy evolution and the
physical state of the ISM. Their combination provides a new chance for
a detailed deconvolution.  While the chemo-dynamical models also
provide element abundances in the hot ICM, their observational
validation is still a problem.

The here applied grid code is still limited to two dimensions because
of the numerically expensive treatment of complex processes and
spatial resolution. It has already been mentioned, that recently also
three-dimensional models were published (e.g.\ \opencite{SG02};
\opencite{HTH02}; \opencite{HTH03}; see also Samland, this
volume). Most approaches are based on N-body simulation and a
multi-phase description of the ISM is achieved by combining a SPH
algorithm, to describe the hot ICM, with a sticky particle scheme, to
describe the cloudy medium. With three-dimensional models infall of
gas clouds or small satellites can be studied as well as the merging
of giant spiral galaxies. This should help to explain e.g. the
question on how and to what extend SF is triggered by such events. 
Nevertheless, one has to keep in mind, that even the two-dimensional
chemo-dynamical investigations present a giant leap with respect to
non-dynamical chemical or purely gas-dynamical studies.

\acknowledgements
S. Harfst is supported by the {\it Deutsche
Forsch\-ungs\-gemein\-schaft (DFG)} under the grant TH-511/2-3. Also,
S. Harfst would like to thank the organisers of WS-ISM and the {\it
Astronomische Gesellschaft} for financial support.

\bibliographystyle{klunamed}           
\bibliography{harfst}

\end{article}
\end{document}